\documentstyle[preprint,aps]{revtex}
\begin{document}

\draft

\title{Theory of Double-Sided Flux Decorations}

\author{M. Cristina Marchetti}
\address{Institute for Theoretical Physics, University of California, Santa
Barbara, Ca 93105}
\address{{\rm and}}
\address{Physics Department, Syracuse University, Syracuse, NY 13244}

\author{David R. Nelson}
\address{Lyman Laboratory of Physics, Harvard University, Cambridge, MA 01238}

\date{\today}

\maketitle

\begin{abstract}
A novel two-sided Bitter decoration technique was recently employed by Yao
et al.
to study the structure of the magnetic vortex array in high-temperature
superconductors. Here we discuss the analysis of such experiments. We show
that two-sided decorations can be used to infer {\it quantitative}
information about the
bulk properties of flux arrays,
and discuss how a least squares analysis of the local density differences
can be used to bring the two sides into registry. Information about
the tilt, compressional and shear moduli of bulk vortex configurations
can be extracted from these measurements.
\end{abstract}
\pacs{PACS: 70.60-w, 74.60Ec}

\widetext
\section{Introduction}

The nature of the ordering of the magnetic flux-line array in the mixed state
of high-temperature superconductors is a topic of much current theoretical and
experimental interest. Most direct measurements of the microscopic structure
of the flux array have been obtained via Bitter decoration experiments at low
fields
\cite{gammel}-\cite{murray}.
The conventional Bitter technique employs small magnetic particles to decorate
the tips of individual vortices as they emerge at one of the sample surfaces.
Since in the high-$T_c$ superconductors the vortices can wander considerably
in the transverse direction as they traverse the sample \cite{seung} and in
addition the
intervortex interaction at the surface differs from that in bulk
\cite{pearl,huse},
it is difficult to unambigously infer bulk properties of flux arrays
from conventional
one-sided decorations that measure two-dimensional correlations of flux-line
tips at
the surface. Quantitative information on the three-dimensional structure of the
vortex
lines as they traverse the sample can be obtained via neutron scattering, but
only very
few
such measurements have been carried out to-date due to their
difficulty and cost. In addition neutron scattering is usually feasible only
at much higher fields than probed by decorations.
Very recently Yao et al. \cite{lieber} used a novel two-sided
decoration technique to study vortex structure in single crystals of BSCCO.
These authors have simultaneously decorated both sides of the sample and
analyzed
how the two-dimensional translational and orientational order of the vortex
array
propagates across its thickness.
Such ``flux transmission spectroscopy'' experiments are likely to become an
important source of insight about vortex matter in the future.
In this paper we discuss the analysis of these experiments and
show how two-sided decorations can be used to infer {\it quantitative}
information
on bulk properties of the flux array. In principle all three bulk
elastic constants of the flux array, the compressional, tilt and shear moduli,
can be extracted from these measurements.

Much of the analysis described below was originally carried out for flux
arrays in a liquid phase \cite{seung},\cite{mcmdis}-\cite{mcmsurf}.
This case may be relevant to the decoration experiments that are
field-cooled below the irreversibility line. Because of long
relaxation times, the observed flux
patterns do not represent the equilibrium configuration of the vortices at the
low temperature where the decoration takes place, but may be better
approximated by the equilibrium configuration
at a higher temperature $T_f$ where the flux flux array falls out of
equilibrium.
The value of $T_f$ is not known, but it is estimated to be
very close to the experimentally observed irreversibility line.
Depending on the field strength, the flux array may
be in a crystalline or liquid-like state when it drops out of equilibrium.
We summarize here results for liquid, hexatic and crystalline vortex arrays.
We also discuss the extent to which configurations on opposite sides of
a sample may be brought into registry by a least square fit of the difference
in the local
vortex density.

We discuss quenched random disorder here only for point pinning and weak
surface disorder in the flux liquid. The effects of bulk point pinning at the
elevated temperatures of the low fields irreversibility line are then
weak because the impurity potential in thick samples is screened out
by thermally induced vortex collisions \cite{ledoussal}.
It should be straightforward, however, to extend much of the analysis
summarized here to other types of bulk and surface pinning in crystalline,
hexatic and liquid phases. Strong surface disorder could certainly obscure
the interpretation of double-sided decorations. If surface pinning
is not a factor, it woud be especially interesting to consider the
effect of {\it correlated} disorder, in the form of columnar pins, either
parallel \cite{drnvv} or splayed \cite{hnv}, which pass completely
through the sample. The key experimental question in this case is whether
vortices always track a single column as they traverse the sample, or if
they hop from column to column. This question plays a particularly
important role in theories of vortex transport in the presence of
splayed defects \cite{hnv}.

\section{Transmission of Density Fluctuations}

Density fluctuations of flux lines in three dimensions
are described by the correlation function of a coarse-grained
areal density field,
\begin{equation}
\label{eq:density}
n({{\bf r}_{\perp}},z)=\sum_{i=1}^N~\delta({{\bf r}_{\perp}}
-{{\bf r}}_i(z)),
\end{equation}
where ${{\bf r}}_i(z)$ is the position of the $i$-th vortex in the $(x,y)$
plane as it wanders along the ${\hat{\bf z}}$
(${\hat{\bf z}}\parallel{\bf H}$) axis.
In a sample of thickness $L$ in the field direction and cross-sectional
area $A$, translational correlations
between the two opposite surfaces of the sample are described by
\begin{eqnarray}
\label{eq:structureL}
n_0 S({q_\perp},L)= & &
   {1\over A}\Big[\overline{<\delta n({{\bf q}_\perp},L)~\delta
n(-{{\bf q}_\perp},0)>}
\nonumber\\
 & & - \overline{<\delta n({{\bf q}_\perp},L)><\delta
n(-{{\bf q}_\perp},0)>}\Big],
\end{eqnarray}
where $\delta n({{\bf q}_\perp},z)=n({{\bf q}_\perp},z)-
n_0A\delta_{{{\bf q}_\perp},{\bf 0}}$ denotes the
fluctuation of the in-plane Fourier transform of the
coarse-grained flux-line density from its equilibrium value $n_0=B/\phi_0$.
A factor of $n_0$ has been extracted in the definition of the structure
factor so that $S({q_\perp},L)\rightarrow 1$ as ${q_\perp}\rightarrow\infty$.
The angular brackets denote a thermal average and the overbar
the average over quenched impurity disorder. The subtracted term
on the right hand side of Eq. \ref{eq:structureL} vanishes in the absence
of quenched disorder.

Almost thirty years ago Pearl \cite{pearl} showed that the interaction
between the tips of straight flux lines at a superconductor-vacuum interface
decays as $1/r_{\perp}$ at large distances, with $r_{\perp}$
the distance between flux tips along the interface. In contrast,
the interaction between flux-line elements in bulk decays exponentially
at large distances.
For this reason Huse argued that at low fields, where the intervortex
separation is large compared to the penetration length,
surface effects may play the dominant role
in determining the magnetic flux patterns seen at the surface
\cite{huse}. The question of the interplay between bulk and surface forces
in determining the vortex structure at the surface was addressed by us
\cite{mcmsurf} with a hydrodynamic
model that incorporates the boundary condition on the flux-lines at the
superconductor-vacuum interface - which is responsible for the $1/r_{\perp}$
interaction - as a surface contribution to the free energy of the flux array,
coupled to the usual bulk free energy. This model neglects all spatial
inhomogeneities
in the $z$ direction other than the presence of the sample boundaries.
The thermal contribution to the structure factor defined in Eq.
\ref{eq:structureL}
was found to be
\begin{equation}
\label{eq:stLres}
S_T({q_\perp},L)=S_{2T}({q_\perp})R({q_\perp},L),
\end{equation}
where $S_{2T}({q_\perp})$ is the two-dimensional structure factor at one
of the two
surfaces and $R({q_\perp},L)$ measures the ``transfer'' of information about
density fluctuations  across
the thickness of the sample. For the hydrodynamic model considered in
Ref. \cite{mcmsurf} this transfer function is given by
\begin{equation}
\label{eq:transltr}
R({q_\perp},L)=  {c_{11}({q_\perp})\xi_{\parallel}\over B_2({q_\perp})
\sinh(L/\xi_{\parallel})
+c_{11}({q_\perp})\xi_{\parallel}\cosh(L/\xi_{\parallel})},
\end{equation}
where $\xi_{\parallel}({q_\perp})=\sqrt{c_{44}({q_\perp})/c_{11}({q_\perp})}
/{q_\perp}$ is the correlation length
describing the decay of in-plane translational order in the $z$ direction
and $c_{11}({q_\perp})$ and $c_{44}({q_\perp})$ are the non-local
compressional and tilt moduli
of the bulk flux array, respectively. The nonlocality of the elastic constants
in the $z$ direction is negligible at low fields compared to the in-plane
variation.
Finally, $B_2({q_\perp})$ is the long wavelength
compressional modulus of a $2d$ liquid of point vortices interacting via
a  $1/{r_{\perp}}$ potential at large distances, so that
$B_2({q_\perp})\approx B^2/4\pi{q_\perp}$ as ${q_\perp}\rightarrow 0$.
As discussed in \cite{mcmsurf}, translational correlations at the surface are
controlled by the $1/r_{\perp}$ surface interaction only for small
wavevectors, such that $B_2({q_\perp})>c_{11}L$, or
${q_\perp}<{q_\perp}^s= {q_\perp} B_2/( L c_{11})$.
For ${q_\perp}>{q_\perp}^s$ the $2d$ surface structure factor is
representative of that of a $2d$ cross-section of bulk and in the hydrodynamic
model it is given by,
\begin{equation}
\label{eq:surfst}
S_{2T}({q_\perp})\approx {n_0 k_BT{q_\perp}^2\over c_{44}({q_\perp})
\xi^{-1}_{\parallel}({q_\perp})}
= {n_0 k_BT\over\sqrt{c_{44}({q_\perp})c_{11}({q_\perp})}}{q_\perp}.
\end{equation}
The transfer function $R({q_\perp},L)$ reduces  then to
\begin{equation}
\label{eq:transfer}
R({q_\perp},L)\approx [\cosh(L/\xi_{\parallel})]^{-1}\approx
2 e^{-L/\xi_{\parallel}({q_\perp})},
\end{equation}
where the second approximate equality holds provided $L>>\xi_{\parallel}$,
i.e., if ${q_\perp}>>{q_\perp}^*=\sqrt{(c_{44}/c_{11})}/L$.
If the elastic constants are calculated from Ginzburg-Landau theory
\cite{fisher},
one finds ${q_\perp}^s\approx{q_\perp}^*\approx 1/L$. On the other hand, there
is evidence
for a strong downward renormalization of the compressional modulus $c_{11}$
from entropic effects at low fields \cite{seung,ledoussal},
as discussed below. In contrast
the tilt modulus is expected to be accurately given by the Ginzburg-Landau
theory.
As a result, ${q_\perp}^*>>1/L$. If the surface bulk modulus $B_2$ is {\it not}
renormalized, then ${q_\perp}^s=B^2/(4\pi Lc_{11})\sim{q_\perp}^* >>1/L$ and
surface effects control the surface translational order up to wavevectors
of order ${q_\perp}^*\sim 10^2/L\sim 5\mu m^{-1}$, where we have used the
parameters
of the Yao et al. experiment with $L=20\mu m$ \cite{lieber}.
On the other hand, we argue below that $B_2$ may also be renormalized downward
by entropic effects due to coupling of the surface tips to line wander
in the bulk.
In this case we expect $B_2\sim c_{11}^R/{q_\perp}$ and
${q_\perp}^s\approx 1/L<<{q_\perp}^*$, so that the long-range surface
interaction only
controls surface translational correlations for wavevectors much smaller
than those probed by the decoration experiments.

The hydrodynamic model is very useful for describing the
long-wavelength properties of the vortex array, and it allows us to incorporate
the nonlocal effects
of the intervortex interaction which are known to be important in flux crystals
over
much of the temperature-field phase diagram. An alternative
more microscopic description
can be obtained via the mapping of flux lines onto the world lines of
two-dimensional
bosons \cite{seung}. In the boson language the correlation length
$\xi_{\parallel}({q_\perp})$ is determined
by the Bogoliubov excitation spectrum $\epsilon({q_\perp})$ of a weakly
interacting superfluid,
according to \cite{seung},
\begin{equation}
\label{eq:bogol}
\xi^{-1}_{\parallel}({q_\perp}) \rightarrow {\epsilon({q_\perp})\over k_BT}
=\sqrt{{n_0V_0\over\tilde{\epsilon}_1}{q_\perp}^2
+\Big({k_BT{q_\perp}^2\over 2\tilde{\epsilon}_1}
\Big)^2},
\end{equation}
where $V_0=\phi_0^2/4\pi=4\pi\lambda^2\epsilon_0$ is the energy scale of the
bare intervortex  interaction and
$\tilde{\epsilon}_1$ is the single vortex tilt energy per unit length,
with where $\lambda$ is the penetration length
in the $ab$ plane.
Although $\tilde{\epsilon}_1\approx (M_\perp/M_z)\epsilon_0\ln\kappa<<
\epsilon_0$ for fields $B>>\phi/\lambda^2$ ($M_\perp/M_z$ is the
effective mass ratio), for $B\simeq\phi/\lambda^2$ (the regime relevant
to decoration experiments) we have $\tilde{\epsilon}_1\approx\epsilon_0$
because of magnetic couplings between the $CuO_2$ planes \cite{ffh}.
The result obtained from the analysis for the boson liquid agrees with
the hydrodynamic result at small ${q_\perp}$, provided we make the
identification
$c_{44}=n_0\tilde{\epsilon}_1$ and $c_{11}=n_0^2V_0$.

Translational order can be quite sensitive to
point disorder, which is present in all experimental samples.
The question of whether the flux patterns seen in decoration experiments
are controlled by quenched disorder or by thermal fluctuations is at present
open. Weak point disorder both in the bulk and at the surface of the sample
can be incorporated in the hydrodynamic model as a random potential with
short-range correlations
coupled to the vortex density \cite{ledoussal}.
Bulk point disorder yields an additive Lorentzian squared correction
to the thermal three-dimensional structure factor. Neglecting surface
effects, the disorder contribution
to the structure factor defined in Eq. \ref{eq:structureL} is given by
\cite{mcmsurf}
\begin{equation}
\label{eq:stdis}
S_D({q_\perp},L)=S_{2D}({q_\perp})(1+L/\xi_\parallel)e^{-L/\xi_\parallel},
\end{equation}
where $S_{2D}({q_\perp})$ is the quenched-disorder contribution to the
two-dimensional structure factor at one of the surfaces,
\begin{equation}
\label{eq:stdissurf}
S_{2D}({q_\perp})=n_0\Delta_B\xi_\parallel\Big({{q_\perp}^2
\xi_{\parallel}\over 2
\tilde{\epsilon}_1}\Big)^2\approx \Delta_B {n_0\over 2 c_{44}}
\Big( {c_{44}\over c_{11}}\Big)^{3/2}{q_\perp},
\end{equation}
and $\Delta_B$ is the correlator of the bulk random impurity potential.
The total structure function is $S({q_\perp},L)=S_T({q_\perp},L)
+S_D({q_\perp},L)$.
The transmittance of translational order across the sample is governed
by the same length scale $\xi_\parallel({q_\perp})$ as in the thermal case
independent of
the strength of the quenched disorder.
Weak surface disorder yields a contribution $S_{2SD}({q_\perp})$
to the two-dimensional
surface structure factor that vanishes as ${q_\perp}^2$ at small wavevectors
and can therefore be distinguished from the other contribuutions
\cite{mcmsurf}.

Double-sided decoration experiments of the type carried out by
Yao et al.
\cite{lieber} can measure both the two-dimensional structure factor
$S_2({q_\perp})$ at one of
the two surfaces, as well the correlations across the thickness of the sample
described by $S({q_\perp},L)$. The transfer function $R({q_\perp},L)$ and
then the
``excitation spectrum'' $\epsilon({q_\perp})/k_BT$
are obtained from
the ratio of these two correlation functions. Since the sample thickness is
known, the slope
of the Bogoliubov spectrum at small ${q_\perp}$ yields a measurement of the
ratio
$\sqrt{c_{11}/c_{44}}$. Yao et al. find
$c_{11}/c_{44}\approx 1.5\times 10^{-4}$ in BSCCO single crystals at $12G$,
a value about four orders of magnitude smaller than predicted from the
Ginzburg-Landau mean field theory \cite{fisher}.
If one assumes that the main contribution to $S_2({q_\perp})$ is the
thermal one
given by Eq. \ref{eq:surfst} and that the flux array falls out of equilibrium
near the irreversibilty temperature $T_{irr}$, i.e., $T_f\approx T_{irr}$,
one can also extract the geometric
mean of the two elastic constants from
the linear slope of $S_2({q_\perp})$ \cite{mcmsurf}.
This gives  $c_{44}\approx 27 G^2$ and
$c_{11}\approx 6\times 10^{-3}G^2$ \cite{lieber}.
The value of $c_{44}$ is essentially equal to $B^2/4\pi$ and is consistent with
$c_{44}\approx n_0\epsilon_0$, provided one uses the value of $\lambda$ at the
irreversibility line \cite{notetilt}.
The expression for the nonlocal tilt modulus obtained from the Ginzburg-Landau
theory can be found in the Appendix of Ref. \cite{fisher}. If we use the
expression for $c_{44}$ that applies at low fields ($a_0\leq\lambda$,
but ${q_\perp}>>\sqrt{M_\perp/M_z}/\lambda$ so that nonlocal effects in
the tilt
modulus can be neglected),
we obtain $c_{44}\approx n_0\epsilon_0/2$, consistent with the
experimental measurement. The experimental value for $c_{11}$ is about four
order of magnitudes smaller than expected on the basis of Ginzburg-Landau
theory, which neglects fluctuation effects.

To obtain an approximate understanding of the strong downward renormalization
of the compressional
modulus (or equivalently of the strength of the intervortex interaction),
we recall that the
Bogoliubov results can be made quantitatively accurate for dilute superfluids,
provided the bare interaction potential is replaced by an effective interaction
or ``$t$-matrix'' defined as the sum of an infinite series of ladder diagrams
\cite{fishhal}.
In a two-dimensional superfluid gas the renormalization of the ${q_\perp}=0$
part of
the intervortex interaction corresponds to the summation of a series in
$1/\ln(1/n_0\lambda^2)$, where $n_0$ is the boson density and $\lambda$
the range of the interaction, and leads to the replacement
\cite{seung,ledoussal}
\begin{eqnarray}
\label{eq:renint}
V_0\rightarrow V_R& &={V_0\over
1+[V_0\tilde{\epsilon}_1/(k_BT)^2]\ln(1/n_0\lambda^2)/4\pi}\nonumber \\
& &\approx {4\pi(k_BT)^2\over\tilde{\epsilon}_1\ln(1/n_0\lambda^2)}.
\end{eqnarray}
The renormalized compressional modulus is then estimated as $c_{11}^R\approx
n_0^2V_R$.
Substituting the material parameters appropriate to the experiments of Ref.
\cite{lieber}, we find $c_{11}^R\sim 2\times 10^{-4}G^2$.
The experiments may not be in the limit of extreme dilution ($n_0\lambda^2<<1$)
required for the second line of Eq. \ref{eq:renint}, so it is not surprising
that this result is
even {\it lower} than the experimental value.

When $n\lambda^2\geq {\cal O}(1)$ one can qualitatively expect an analogous
downward renormalization to arise from entropic
contributions to the free energy
from vortex-line braiding \cite{fnf}.
In this dense limit each vortex line spends
a certain ``time'', i.e., length along the $z$ axis,
in the tube or ``cage'' of radius $a_0$ provided
by the repulsive interaction with its six neighbors. Flux lines
wander within this cage until they
escape to one
of the approximately six neighboring cages.
Collisions reduce the entropy of the
interacting flux array relative to that of the noninteracting system.
Escape events, which yield flux-line braiding, {\it increase}, however.
the entropy,
similar to the discussion of interstitial wandering in Ref. \cite{fnf}.
In the flux-line liquid, where escapes are frequent, the reduction in
entropy due to collisions, or unsuccessful escapes, is a small correction.
Each escape increases the entropy per vortex by $k_B\ln q$, with $q$ an
effective coordination number describing the different
directions in which a vortex can hop. The average distance $l_z$
between hops among lattice sites is given by $D_0l_z\sim a_0^2$,
with $D_0=k_BT/\tilde{\epsilon}_1$ the vortex ``diffusion constant''
along $\hat{\bf z}$,
or $l_z\approx \tilde{\epsilon}_1a_0^2/k_BT$ \cite{seung}.
In a sample of thickness $L$ the total number of jumps
is of order $L/l_z$ and the corresponding entropic contribution to
the Gibbs free energy per unit volume of the vortex array is
$g_{ent}\approx -{N\over AL}k_BT{L\over l_z}\ln q=
-{(k_BT)^2\over\tilde{\epsilon}_1}n^2\ln q$.
The total Gibbs free energy per unit volume can be written as
$g(n)\approx -n{\phi_0\over 4 \pi}(H-H_{c1})+g_{int}(n)+g_{ent}(n)$,
where $g_{int}\approx\epsilon_0n^2\lambda^2$ is the contribution from
intervortex interactions. Upon expanding about the minimum density
to otain $c_{11}=(d^2g/dn^2)|_{n=n_0}$, we see that the entropic contribution
partially
cancels the large contribution from interactions,
consistent with experimental observations.
The entropic and energetic contributions to $c_{11}$ are comparable when
$T\approx\sqrt{\epsilon_0\tilde{\epsilon}_1}\lambda$, which is comparable
to the
melting temperature of the Abrikosov flux lattice in this field regime
\cite{fnf}.

An analogous mechanism could lead to a strong downward renormalization
of the surface interaction. This is because the flux tips at the sample surface
are not true point vortices, but are connected to the flux lines in
the bulk. Braiding effects of the type described above within a
surface layer of thickness $\xi_z({q_\perp})\sim\sqrt{c_{44}/c_{11}}/{q_\perp}$
will increase the surface entropy of the flux-tips, yielding
a contribution
$g^s_{ent}\approx -k_BT n (\xi_z/l_z)\ln q\approx -{(k_BT)^2\over
\sqrt{\tilde{\epsilon}_1V_0}{q_\perp}}n^{3/2}\ln q$
to the free energy per unit area, where we have used $c_{44}\sim
n\tilde{\epsilon}_1$ and $c_{11}\sim n^2V_0$.
The corresponding free energy from surface interaction among
the flux tips is $g_{int}\sim n^2{\phi_0^2\over 4\pi{q_\perp}}$.
Provided $\tilde{\epsilon}_1\sim\epsilon_0$, as is appropriate for
this low field regime, the two contributions to the energy
(and hence to $B_2^R({q_\perp})$) are again comparable near the melting
temperature.

The finite range of the intervortex interaction can be incorporated in the
derivation of the Bogoliubov spectrum, which is then given by Eq.
\ref{eq:bogol}, with
$V_0\rightarrow V({q_\perp})=V_0/(1+{q_\perp}^2\lambda^2)$
\cite{ledoussal,leo}.
The second term in Eq. \ref{eq:bogol}
is unchanged
since it represents the ``kinetic energy'' contribution to the spectrum,
which is unrenormalized due to Galileian invariance of the equivalent
boson problem.
The full wavevector-dependent
interaction $V({q_\perp})$ is again renormalized by resumming an infinite
series of ladder
diagrams, which leads to an integral equation for the effective $t$-matrix
at finite wavevector \cite{mcmint}. Upon neglecting corrections logarithmic
in the wavevector, we find that
the screening length $\lambda$ is not renormalized and one obtains
$V_R({q_\perp})=V_R/(1+{q_\perp}^2\lambda^2)$.
The Bogoliubov spectrum can then be rewritten
in a suggestive form that interpolates between the boson result and the
hydrodynamic
description as
\begin{equation}
\label{eq:renbog}
\xi^{-1}_{\parallel}({q_\perp}) \rightarrow {\epsilon({q_\perp})\over k_BT}
=\sqrt{{c^R_{11}({q_\perp})\over c_{44}}{q_\perp}^2
+\Big({k_BT{q_\perp}^2\over 2\tilde{\epsilon}_1}
\Big)^2},
\end{equation}
where we have identified the renormalized local compressional modulus as
$c^R_{11}({q_\perp})=n_0^2V_R({q_\perp})$. The nonlocality of the tilt
modulus is not
important for the low fields of interest here \cite {fisher},
as confirmed by the experimental
finding that $c_{44}\approx n_0\tilde{\epsilon}_1$.
The renormalized Bogoliubov spectrum given in Eq. \ref{eq:renbog} is
shown in Fig. 1.

The Bogoliubov spectrum is not expected to be quantitatively accurate for
dense superfluids.
In this regime the theory can be improved following Feynman and approximating
this
spectrum by
\begin{equation}
\label{eq:roton}
{\epsilon({q_\perp})\over k_BT}={k_BT{q_\perp}^2\over
2n_0\tilde{\epsilon}_1 S_2({q_\perp})},
\end{equation}
where $S_2({q_\perp})$ is the structure factor of a two-dimensional cross
section of a
dense vortex liquid. Its thermal contribution in hydrodynamic theory
is given in Eq. \ref{eq:surfst}.
For more realistic functions $S_2({q_\perp})$, this formula leads
to a ``roton'' minimum in the excitation spectrum
at ${q_\perp}\approx k_{BZ}=\sqrt{4\pi n_0}$, at approximately the position
of the first maximum in $S_2({q_\perp})$.
This ``roton'' minimum has been observed in the experiments
\cite{lieber}.

\section{Translational and Rotational Registry}

A source of uncertainty arises in the experiments from the difficulty in
matching the $(x,y)$ locations of vortices
being imaged on the two sides of the sample.
This positional uncertainty can be decreased when the sample contains localized
defects that run all the way across the sample, such as grain boundaries,
since these can provide a common reference frame on the two sides
\cite{lieber}.
Correlations on different length scales are in general affected differently
by this mismatch. To quantify this effect for a given sample we define
the $rms$ density fluctuations arising from a translational mismatch
$d$ and an orientational mismatch $\phi$
averaged over the area $A$ of the sample,
as
\begin{equation}
\label{eq:rms}
\Delta(d,\phi,L)=\int {d^2{{\bf r}_{\perp}}\over A}
<[\delta n({\cal R}_\phi\cdot{{\bf r}_{\perp}}
+{\bf d},L)-\delta n({{\bf r}_{\perp}},0)]^2>,
\end{equation}
where ${\cal R}_\phi$ is a two-dimensional rotation matrix and we
neglect effects due to quenched random disorder.
We expect $\Delta(d,\phi,L)$ to be a minimum when the patterns on the
two sides are defined relative to $(x,y)$ coordinate systems with
a common origin and orientation. Minimizing $\Delta$ with respect to
$d$ and $\phi$ using experimental data could be used to
bring these coordinate systems
into registry even in the absence of identifying features such as
grain boundaries which penetrate across the entire crystal.
Upon introducing the Fourier components of the density, Eq. \ref{eq:rms}
can be rewritten using the single-pole approximation in $q$-space for the
structure factor,
\begin{equation}
\label{eq:structure}
\hat{S}({\bf q}_\perp,q_z)=<|\delta\hat{n}({\bf q}_\perp,q_z)|^2>=
{n_0^2k_BT q_\perp^2/c_{44}\over q_z^2+[\epsilon(q_\perp)/k_BT]^2},
\end{equation}
as,
\begin{eqnarray}
\label{eq:rmsans}
\Delta(d,\phi,L)={n_0^2k_BT\over 2\pi c_{44}}\int_0^{k_{BZ}} d{q_\perp}
& & {q_\perp}^3
{k_BT\over\epsilon_R({q_\perp})}\nonumber \\
& & \times\Big[1-J_0({q_\perp} d){J_1(2Rk_{BZ}\sin(\phi/2))\over
Rk_{BZ}\sin(\phi/2)}
e^{-L\epsilon_R({q_\perp})/k_BT}\Big],
\end{eqnarray}
where $J_0(x)$ and $J_1(x)$ are Bessel functions and $R$ denotes the
linear dimensions of the sample in the $ab$ plane, with $A=\pi R^2$.
The integral on the right hand side of
Eq. \ref{eq:rmsans} diverges and is cutoff by a circular
Brillouin zone, $k_{BZ}=\sqrt{4\pi n_0}$. We have evaluated the
dimensionless quantity
$\tilde{\Delta}(d,\phi,L)=
[\Delta(d,\phi,L)-\Delta(0,0,L)]/\Delta(0,0,L)$
using the hydrodynamic approximation
$\epsilon_R(q_\perp)/k_BT=q_\perp\sqrt{c_{11}^R(q_\perp)/c_{44}}$.
The function $\tilde{\Delta}(d,\phi,L)$ is shown in Figs. 2
as a function of both the angle $\phi$ and the translation $d$
for a few values of the sample
thickness. The function $\Delta(d,\phi,L)$ has a parabolic minimum
at $d=0$, $\phi=0$, according to
\begin{equation}
\label{eq:parab}
\Delta(d,\phi,L)\approx \Delta(0,0,L)\bigg\{1+{1\over 2}\alpha(L)
\Big[d^2k_{BZ}^2/2+R^2k_{BZ}^2\sin^2(\phi/2)\Big]\bigg\},
\end{equation}
where the dimensionless curvature $\alpha(L)$ is given by
\begin{equation}
\label{eq:curv}
\alpha(L)={\int_0^1 dx x^2\sqrt{1+x^2\lambda^2k_{BZ}^2}
e^{-L^*k_{BZ}{x\over\sqrt{1+x^2\lambda^2k_{BZ}^2}}}\over
\int_0^1 dx x^2\sqrt{1+x^2\lambda^2k_{BZ}^2}
\bigg[1-e^{-L^*k_{BZ}{x\over\sqrt{1+x^2\lambda^2k_{BZ}^2}}}\bigg]},
\end{equation}
and $L^*=L\sqrt{c_{11}(q_\perp=0)/c_{44}}$.
At low density, corresponding to $\lambda k_{BZ}<<1$, the behavior of
the curvature is controlled by $L^*k_{BZ}$, with
$\alpha\sim 1/(L^*k_{BZ})$ for $L^*k_{BZ}<<1$ and
$\alpha\sim 1/(L^*k_{BZ})^5$ for $L^*k_{BZ}>>1$.
At high density,  corresponding to $\lambda k_{BZ}>>1$, the
relevant length scale is $L^/\lambda$ and
$\alpha\sim \lambda/L^*$ for $L^*<<\lambda$ and
$\alpha\sim 1/(L^*k_{BZ})^5$ for $L^*>>\lambda$.

\section{Transmission of Orientational Order}

{}From the analysis of double-sided decorations one can also study the
propagation of orientational order across the sample.
Orientational order is much less sensitive to point pinning \cite{chudnovsky}.
It is measured by correlations  in the bond-orientational
order parameter $\psi_6({{\bf r}})=e^{6i\theta({{\bf r}})}$, where
$\theta({{\bf r}})$
is the bond-angle field. The corresponding angular correlation across
the sample thickness
is
\begin{eqnarray}
\label{eq:bond}
G_H({{\bf r}_{\perp}},L)& &=<e^{6i[\theta({{\bf r}_{\perp}},L)-
\theta(0,0)]}>\nonumber \\
& & \approx \exp[-18<[\theta({{\bf r}_{\perp}},L)-\theta(0,0)]^2>].
\end{eqnarray}
The decay of bond-orientational order in a hexatic flux liquid was discussed in
Ref. \cite{mcmdis} in the hydrodynamic limit.
In a bulk sample, ignoring boundary conditions and surface effects,
the in-plane Fourier transform of the thermal part of the correlation
function of the bond-orientational order parameter
was found to be given by
\begin{equation}
\label{eq:hex}
G_H({q_\perp},L)=<|\psi_6|^2>\Big[A\delta_{{{\bf q}_\perp},{\bf 0}}
+G_{H2}({q_\perp})
e^{-L/\xi_H({q_\perp})}\Big],
\end{equation}
with
\begin{equation}
\label{eq:sthex}
G_{H2}({q_\perp})=
{9k_BT\over K_A^z}\xi_H({q_\perp})={9k_BT\over\sqrt{K_A^zK_A^\perp}}
{1\over{q_\perp}}.
\end{equation}
Here $\xi_H({q_\perp})=\sqrt{K_A^z/K_A^\perp}/{q_\perp}$
is the correlation length
governing the transmittance of hexatic order across an hexatic flux-line
liquid and
$K_A^z$ and$ K_A^\perp$ are the hexatic stiffnesses.

In a superconducting slab of finite thickness $L$ we use free boundary
conditions on the bond-angle field at the surface to find that Eq.
\ref{eq:hex} is replaced by,
\begin{equation}
\label{eq:bondfin}
G_H({q_\perp},L)=<|\psi_6|^2>\Big[A\delta_{{{\bf q}_\perp},{\bf 0}}+
G_{H2}({q_\perp})R_H({q_\perp},L)\Big],
\end{equation}
with $G_{H2}({q_\perp})={k_BT\over K_A^z}\xi_H({q_\perp})\coth(L/\xi_H)$
and $R_H({q_\perp},L)=[\cosh(L/\xi_H)]^{-1}$.

In a flux lattice with long-range crystalline order the hexatic order parameter
is not independent, but is simply related to the curl of the elastic
diplacement field, $\theta={1\over 2}\hat{\bf z}\cdot (\vec{\nabla}\times
\vec{u})$.
The correlation function of the bond-angle field in an infinite sample
then follows immediately, with the result,
\begin{equation}
\label{eq:hexlatt}
G_H^L({q_\perp},L)=<|\psi_6|^2>\Big[A\delta_{{{\bf q}_\perp},{\bf 0}}
+G^L_{H2}({q_\perp})
e^{-L/\xi_H^L({q_\perp})}\Big],
\end{equation}
where $\xi_H^L({q_\perp})=\sqrt{c_{44}\over c_{66}}/{q_\perp}$
is the correlation length
governing transmittance of hexatic order across a flux lattice, and
\begin{equation}
\label{eq:sthexlat}
G_{H2}^L({q_\perp})={9k_BT\over 4c_{44}\xi_H^L({q_\perp})}
={9k_BT\over 4\sqrt{c_{66}c_{44}}}{q_\perp}.
\end{equation}
The corresponding expressions in a finite-thickness sample with free
boundary condition on the bond-angle field are modified with the same
finite-size functions of $L/\xi_H^L({q_\perp})$ as in the case of the hexatic
flux liquid.
This result shows that by measuring the correlation of bond order across the
sample, as well as $G_{H2}^L({q_\perp})$ at one of the surfaces, one can
infer the
value of the tilt and shear moduli. In addition, Eqs.  \ref{eq:sthex} and
\ref{eq:sthexlat} show that surface bond-orientational order decays as
$1/{q_\perp}$ in a hexatic liquid, but grows as ${q_\perp}$ in a lattice,
providing a
further mean to distinguish hexatic and crystalline order in the vortex array.

Bond order decays exponentially in flux liquids which are isotropic in a
plane perpendicular to the field direction. The results in this case are
similar to Eq. \ref{eq:hex}, except that the delta-function term is
absent and $\lim_{{q_\perp}\rightarrow 0}\xi_H({q_\perp})$ is finite and
equal to the
hexatic correlation length along the $z$ direction.

\vskip .2in
It is a pleasure to acknowledge helpful conversations with C. Lieber,
H. Dai, S. Yoon and Leon Balents.
This work was supported by the National Science Foundation at Syracuse through
Grants DMR-9112330 and DMR-9217284, at Harvard primarily by the
MRSEC Program through Grant DMR-9400396 and through Grant
DMR-9417047,
and at the ITP of the University of California in Santa Barbara
through Grant PHY-8904035.


\newpage
\centerline{Figure Captions}
\noindent
Fig.~1. The renormalized Bogoliubov spectrum given in Eq.
(\protect\ref{eq:renbog})
as a function of wavevector for $n_0\lambda^2=0.2,0.5,1$.

\vskip 1.0truecm
\noindent
Fig.~2. The spatially averaged mismatch function $\tilde{\Delta}(d,\phi,L)$
is shown (a) at $\phi=0$ as a function of $d$ and (b) at $d=0$ as a
function of $\phi$, for three values of $L$. Note that the sample thickness
only enters in the dimensionless combination
$L k_{BZ}\sqrt{c_{11}(q_\perp=0)/c_{44}}$.
We have used $B=12G$, $c_{11}(q_\perp=0)/c_{44}=1.5\times 10^{-4}$ and
$R=0.2 mm$.


\begin{references}

\bibitem{gammel}
P.L. Gammel, D.J. Bishop, G.J. Dolan, J.R. Kwo,
C.A. Murray, L.F. Schneemeyer, and J.V. Waszczak, Phys. Rev. Lett.
{\bf 59}, 2592 (1987).

\bibitem{dolan}
G.J. Dolan, G.V. Chandrashekhar, T.R. Dinger,
C. Feild and F. Holtzberg, Phys. Rev. Lett. {\bf 62}, 827 (1989).

\bibitem{murray}
C.A. Murray, P.L. Gammel,
D.J. Bishop, D.B. Mitzi, and A. Kapitulnik, Phys. Rev. Lett. {\bf 64},
2312 (1990); D.G. Grier, C.A. Murray, C.A. Bolle, P.L. Gammel,
D.J. Bishop, D.B. Mitzi, and A. Kapitulnik, Phys. Rev. Lett.
{\bf 66}, 2270 (1991).

\bibitem{seung}
D.R. Nelson and S. Seung, Phys. Rev. B{\bf 39}, 9153 (1989).

\bibitem{pearl}
J. Pearl, J. App. Phys. {\bf 37}, 4139 (1966).

\bibitem{huse}
D. A. Huse, Phys. Rev. B{\bf 46}, 8621 (1992).

\bibitem{lieber}
Z. Yao, S. Yoon, H. Dai, S. Fan and C.M. Lieber, Nature {\bf 371},
777 (1994), and to be published.

\bibitem{mcmdis}
M.C. Marchetti and D.R. Nelson, Phys. Rev. B{\bf 41}, 1910 (1990).

\bibitem{ledoussal}
D.R. Nelson and P. LeDoussal, Phys. Rev. B{\bf 42}, 10113 (1990);
see footnote \protect\cite{fisher}.

\bibitem{leo}
L. Radzihovsky and E. Frey, Phys. Rev. B{\bf 48}, 10357 (1993).

\bibitem{mcmsurf}
M.C. Marchetti and D.R. Nelson, Phys. Rev. B{\bf 47}, 12214 (1993).

\bibitem{drnvv}
D.R. Nelson and V.M. Vinokur, Phys. Rev. B{\bf 48}, 13060 (1993).

\bibitem{hnv}
T. Hwa, P. LeDoussal, D.R. Nelson and V.M. Vinokur, Phys. Rev. Lett.
{\bf 71}, 3545 (1993).

\bibitem{ffh}
D.S. Fisher, M.P.A. Fisher and D. Huse, Phys. Rev. B{\bf 43}, 130 (1991).

\bibitem{fisher}
D.S. Fisher, in {\it Phenomenology and Applications of High-Temperature
Superconductors}, K.S. Bedell, ed. (Addison-Wesley, Reading, MA, 1991),
pp. 321-324.

\bibitem{fishhal}
V.N. Popov, {\it Functional Integrals in Quantum Field Theory and
Statistical Physics} (D. Reidel Pub. Co., Dordrecht, 1983);
D.S. Fisher and P.C. Hohenberg, Phys. Rev. B{\bf 37}, 4936 (1988).

\bibitem{notetilt}
The lack of renormalization of $c_{44}$ at low fields corresponds to the
lack of renormalization of the superfluid density of a weakly interacting
gas of bosons that arises from Galileian invariance.

\bibitem{fnf}
E. Frey, D.R. Nelson and D.S. Fisher, Phys. Rev. B{\bf 49}, 9723 (1994).

\bibitem{mcmint}
M.C. Marchetti and D.R. Nelson, unpublished.

\bibitem{chudnovsky}
E.M. Chudnovsky, Phys. Rev. B{\bf 43}, 7831 (1991).


\end{references}
\end{document}